# Crosstalk based Fine-Grained Reconfiguration Techniques for Polymorphic Circuits

Naveen Kumar Macha, Sandeep Geedipally, Bhavana Repalle, Md Arif Iqbal, Wafi Danesh, Mostafizur Rahman
Department of Computer Science & Electrical Engineering, University of Missouri Kansas City, MO, USA
E-mail: nmhw9@mail.umkc.edu, rahmanmo@umkc.edu

*Abstract*— Truly polymorphic circuits, whose functionality/circuit behavior can be altered using a control variable, can provide tremendous benefits in multi-functional system design and resource sharing. For secure and fault tolerant hardware designs these can be crucial as well. Polymorphic circuits work in literature so far either rely on environmental parameters such as temperature, variation etc. or on special devices such as ambipolar FET, configurable magnetic devices, etc., that often result in inefficiencies in performance and/or realization. In this paper, we introduce a novel polymorphic circuit design approach where deterministic interference between nano-metal lines is leveraged for logic computing and configuration. For computing, the proposed approach relies on nano-metal lines, their interference and commonly used FETs, and for polymorphism, it requires only an extra metal line that carries the control signal. In this paper, we show a wide range of crosstalk polymorphic (CT-P) logic gates and their evaluation results. We also show an example of a large circuit that performs both the functionalities of multiplier and sorter depending on the configuration signal. Our benchmarking results are presented in this paper. For CT-P, the transistor count was found to be significantly less compared to other existing approaches, ranging from 25% to 83%. For example, CT-P AOI21-OA21 cell show 83%, 85% and 50% transistor count reduction, and Multiplier-Sorter circuit show 40%, 36% and 28% transistor count reduction with respect to CMOS, genetically evolved, and ambipolar transistor based polymorphic circuits respectively.

## I. INTRODUCTION

Polymorphic circuits have found use in a myriad of application areas, ranging from enhanced functionality to resource sharing, fault tolerance, and cybersecurity. In literature, many attempts [2-12] for circuit polymorphism can be found. A simple approach is to have multiple functional blocks, which are selected using a multiplexer. A variation of this approach superimposes functionalities on CMOS circuits [2]. These design approaches face key limitations such as circuit overhead and design complexity. In another category, polymorphic circuits are designed using genetic algorithms [3][4]. In this approach, the circuit behavior is morphed using different control variables such as temperature, power supply voltage, light, control signal etc. These type of circuits are strongly dependent on conditions and technology under which they are evolved and suffer from lack of general circuit topologies, slow and unreliable output responses, higher power consumption etc. More recently, polymorphic circuits are also designed using emerging tunable polarity transistors presented in [7-8]. They are based on ambipolar property achievable in silicon-nanowires [7-8], carbon nanotubes [9], organic layered transistors [10] etc. Though polymorphic complementary-style circuits [7][8] using these reconfigurable p-type/n-type transistors have been designed, they require complex device engineering and additional circuitry, also, the circuits are not very compact. The other approaches using emerging spintronic devices were also proposed [11], but they rely on complex information encoding scheme through spin-polarized currents and bipolar voltages etc.

In contrast to these approaches, we propose a novel solution to achieve multifunctional circuits in an efficient manner. In this approach, we embrace the increasing crosstalk signal interference at advancing technology nodes and astutely engineer it to a logic principle [1]. For operation, the transition of signals on input metal lines (including polymorphic control signal) called as aggressor nets induce a resultant summation charge on output metal line called as victim net through capacitive couplings. This induced signal serves as an intermediate signal to control thresholding devices like pass-transistor or an inverter to get desired logic output. To achieve polymorphic behavior, the victim net is influenced/biased by a control aggressor, which switches the circuit behavior to a different logic type. We demonstrate the intrinsic multifunctional ability of crosstalk circuits by showing various polymorphic circuit implementations. The circuits implemented are NAND2/AND2 to NOR2/OR2, AOI21/AO21 to OAI21/OA21, NAND3/AND3 to OAI21/OA21, NAND3/AND3 to AOI21/AO21, NOR3/OR3 to OAI21/OA21, and NOR3/OR3 to AOI21/AO21. These basic polymorphic cells are very compact and use only 3 to 5 transistors. We also demonstrate a larger circuit i.e. Multiplier-Sorter circuit using CT-P gates. Due to the polymorphic nature of crosstalk gates, the same circuit can be switched between multiplier or sorter operation depending on control aggressor value (low or high).

The rest of the paper is organized as follows. Section II describes the crosstalk (CT) computing fabric and the implementation of fundamental logic gates. Section III presents a wide range of basic and complex polymorphic logic gates implemented in CT-P logic. A cascaded circuit example and subsequent discussion on signal integrity are also presented in this section. Section IV compares and benchmarks CT-P logic with other polymorphic circuits available in the literature. Finally, section V concludes the paper.

## II. CROSSTALK COMPUTING FABRIC

The logic computation in crosstalk computing fabric happens in metals lines, coupled with accurate control and reconstruction of signals in transistors. This is depicted in (Fig.1), where the capacitive interference of the signals for logic computation takes place in metal layer 2 (aggressor and victim nets) and the bottom layer is for transistors performing two functions: periodically controlling the floating victim nodes and re-boosting the signals using inverters. The metal layer 1 is used for power rails and local routing. The inset figure illustrates the aggressor-victim scenario of crosstalk-logic, the transition of the signals on two adjacent aggressor metal lines (*Ag1* and *Ag2*) induces a resultant summation charge/voltage on victim metal line (*Vic*) through capacitive coupling. Since this phenomenon follows the charge conservation principle, the victim node voltage is deterministic in nature. Therefore, it can be stated that the signal induced on the victim net possesses the information about signals on two aggressor nets, and its magnitude depends upon the coupling strength between the aggressors and victim net. This coupling capacitance is inversely proportional to the distance of separation of metal lines and directly proportional to the relative permittivity of the dielectric and lateral area of metal lines (which is length *times* vertical thickness of metal lines). Tuning the coupling capacitance values using the variables mentioned above provides the engineering freedom to tailor the induced summation signal to the specific logic implementation or as an intermediate control signal for further circuitry. Therefore, the geometrical arrangement and dielectric choice of aggressor and victim metal lines in CT-logic are according to the coupling requirement for specific logic. For example, OR gate requires stronger coupling than AND gate, which can be achieved by tuning the dimensions and high-k dielectric material choices.

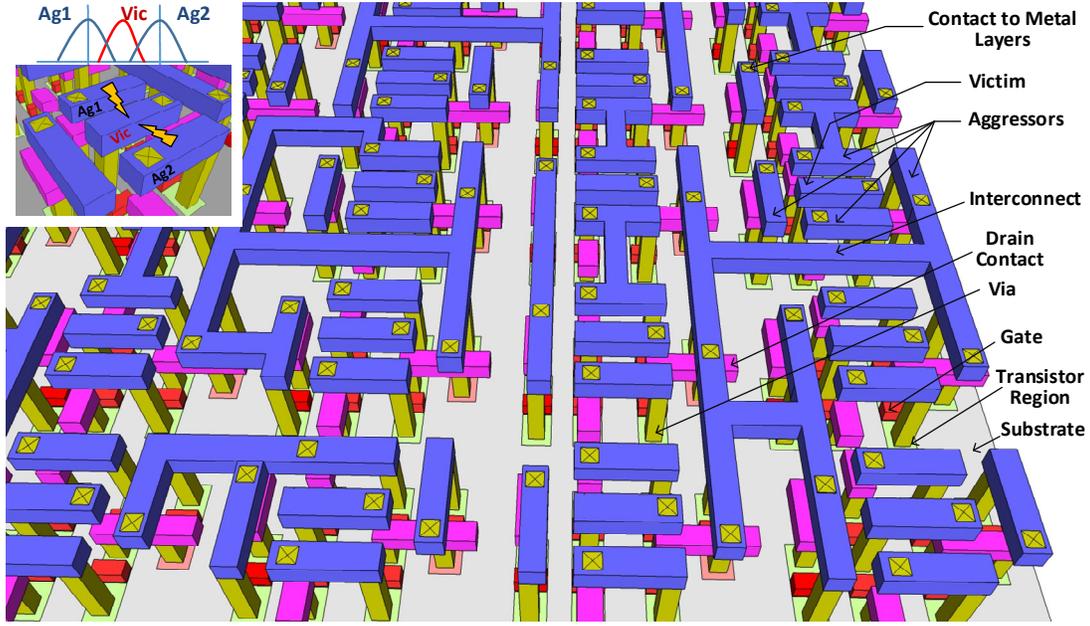

Fig.1. Abstract view of the Crosstalk computing fabric.

*A. Fundamental Logic Gates*

We have introduced the crosstalk computing concept in [1]. The CT-logic can implement efficiently both linear logic functions (e.g., AND, OR etc.), non-linear logic functions (e.g. XOR). Moreover, it offers compact and effective reconfiguration between these functions, both linear to linear, and linear to non-linear are possible. In this paper, we demonstrate only linear logic functions. Fig.2(i) and 2(ii) show the NAND and NOR circuits in which input aggressor nets (*A* and *B* acting as *Ag1* and *Ag2*) are coupled to victim net (*Vi*) through coupling capacitances $C_{ND}$ and $C_{NR}$ respectively. A discharge transistor driven by *'Dis'* signal and an inverter are connected to *'Vi'* net as shown in the figure. The CT-logic operates in two states, logic evaluation state and discharge state (DS). During logic evaluation state, the rise transitions on aggressor nets induce a proportional linear summation voltage on *Vi* (through couplings) which is connected to a CMOS inverter acting as a threshold function. During discharge state (enabled by *Dis* signal) floating victim node is shorted to ground through discharge transistor, which ensures correct logic operation during next logic evaluation state by clearing off the value from the previous logic operation. The simulation response of the designed NAND and NOR gates are shown in Fig.2(iii), the first panel shows the discharge pulse (*Dis*), the second panel shows two input signals (*A* and *B*) with 00 to 11 combinations given through successive evaluation stages (when *Dis*=0). Third and fourth panels show the output response of NAND and NOR gates respectively. It is to be noted that, as victim node is discharged to ground in every DS (*Dis*=1) the outputs of these gates are logic high. The operation of CT logic gates would be represented functionally using a crosstalk-margin function $CT_M(C)$, which specifies that the inverter of the CT-logic gate flips its state only when victim node sees the input transitions through the total coupling greater than or equal to C. For example, as shown in the Fig.2(i), NAND CT-margin function is $CT_M(2C_{ND})$, which states that inverter flips the state only when victim node sees the input transitions through total coupling greater than or equal to $2C_{ND}$, i.e. when both inputs are high. Similarly, for NOR gate (Fig.2(ii)) the CT-margin function is $CT_M(C_{NR})$, which means the transition of any one of the aggressor is sufficient to flip the inverter, thus evaluates to NOR behavior.

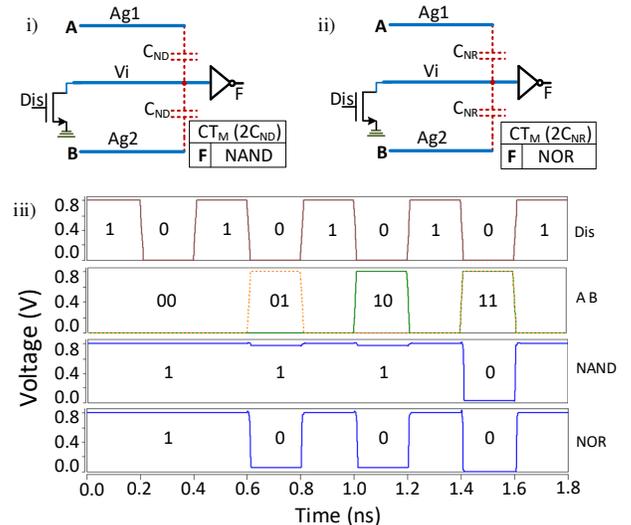

Fig.2. CT basic gates, i) CT-NAND gate, ii) CT-NOR gate, iii) Simulation results

CT-logic can implement complex logic functions efficiently in a single stage, which is discussed next. Fig.3(i) and 3(ii) show AOI21 and OAI21 cells. Logic expression of AOI21, *(AB+C)'*, evaluates to 0 when either *AB* or *C*, or both are 1. That means the output is biased towards the input *C* i.e., irrespective of *A* and *B* values, the output is 1 when *C* is 1. Therefore, in Fig.3(i), input *C* has the coupling $2C_{AO}$, whereas, *A* and *B* have $C_{AO}$ capacitance. The margin function for this gate is $CT_M(2C_{AO})$. The response of the circuit is shown in Fig.3(iii), the first panel shows *Dis* pulse, the second shows the three input signals (*A*, *B* and *C*) feeding all combinations from 000 to 111 in successive logic evaluation states. The third panel shows the response of the AOI21 circuit for the corresponding inputs above, satisfying the logic. Similarly, for OAI21 function, *((A+B)C)'*, the output is biased towards input *C* i.e., for output to be 1, *C* should be 1 along with *A+B*. Therefore, *C* receives $2C_{OA}$, while both *A* and *B* receive $C_{OA}$ each. So the margin function now becomes $CT_M(3C_{AO})$. The fourth panel in Fig.3(iii) shows the response of OAI21 circuit for all input combinations (000 to 111).

From the above circuit implementation and their logic nature, the CT-logic gates are categorized into two types, homogeneous and heterogeneous logic gates. In homogeneous logic gates inputs are coupled equally to the victim net (e.g., NAND and NOR), because the logic behavior is unbiased towards any particular input. With heterogeneous logic gates inputs are coupled unequal to the victim net (e.g., AOI21 and OAI21), because the logic behavior is biased towards certain inputs, as seen with the biased inputs receiving the higher coupling.

### III. CROSS-TALK POLYMORPHIC LOGIC GATES

The polymorphic logic gates exhibit multiple logic behaviors by altering a control variable, as a result, increases the logic expressibility of a circuit. The CT-Polymorphic (CT-P) gates presented in this paper switch the logic behavior by using an additional control aggressor. The reconfigurability is shown between following logics: homogeneous to homogeneous logic type (e.g., AND to OR); heterogeneous to heterogeneous logic type (e.g., AO21 to OA21); and homogeneous to heterogeneous logic type (AND to AO2, AND to OA21, OR to AO21, and OR to OA21). Fig.4(i) shows the CT-P AND-OR circuit and its response graph. As shown in the circuit diagram, inputs (A and B) and control aggressor (*Ct*) has the same coupling $C_{PA}$ ( the coupling capacitance values are detailed in Table.1). $F_1$ stage in Fig.4(i) gives inverting function (NAND/NOR) response and $F$ stage gives non-inverting function (AND/OR). A table adjacent to circuit diagram lists the margin function and the circuit operating modes. The margin function for AND-OR cell is $CT_M(2C_{PA})$. When control Ct=0 it operates as AND, whereas, when Ct=1 the Ct aggressor (*Ag3*) augments an extra charge through the coupling capacitance $C_{PA}$, hence, following the function $CT_M(2C_{PA})$ the cell is now biased to operate as an OR gate, therefore, the transition of either *A* or *B* is now sufficient to flip the inverter. The same response can be observed in the simulation plots shown in Fig.4(i), the first panel shows the discharge (*Dis*) and control (*Ct*) signals, 2nd panel shows the input combinations fed through *A* and *B*, and 3rd panel shows the response at stage *F*. It can be observed that the circuit responds as AND when Ct=0 for first four input combinations (00 to 11), whereas, it responds as OR when Ct=1 during next four input combinations (00 to 11).

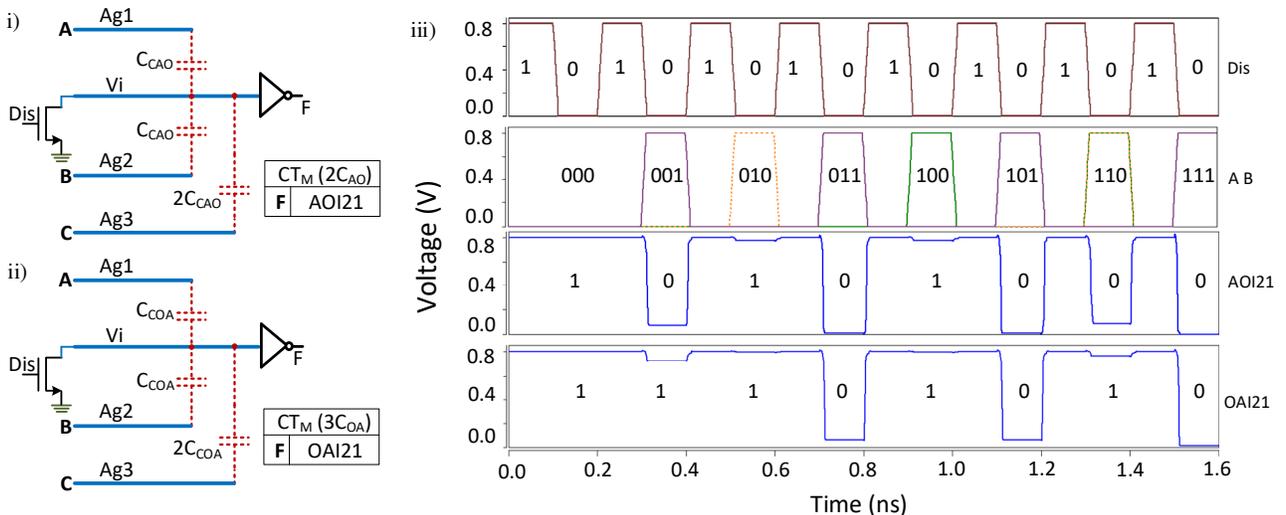

Fig.3. CT Complex gates, i) CT-AOI21, ii) CT-OAI21, iii) Simulation results

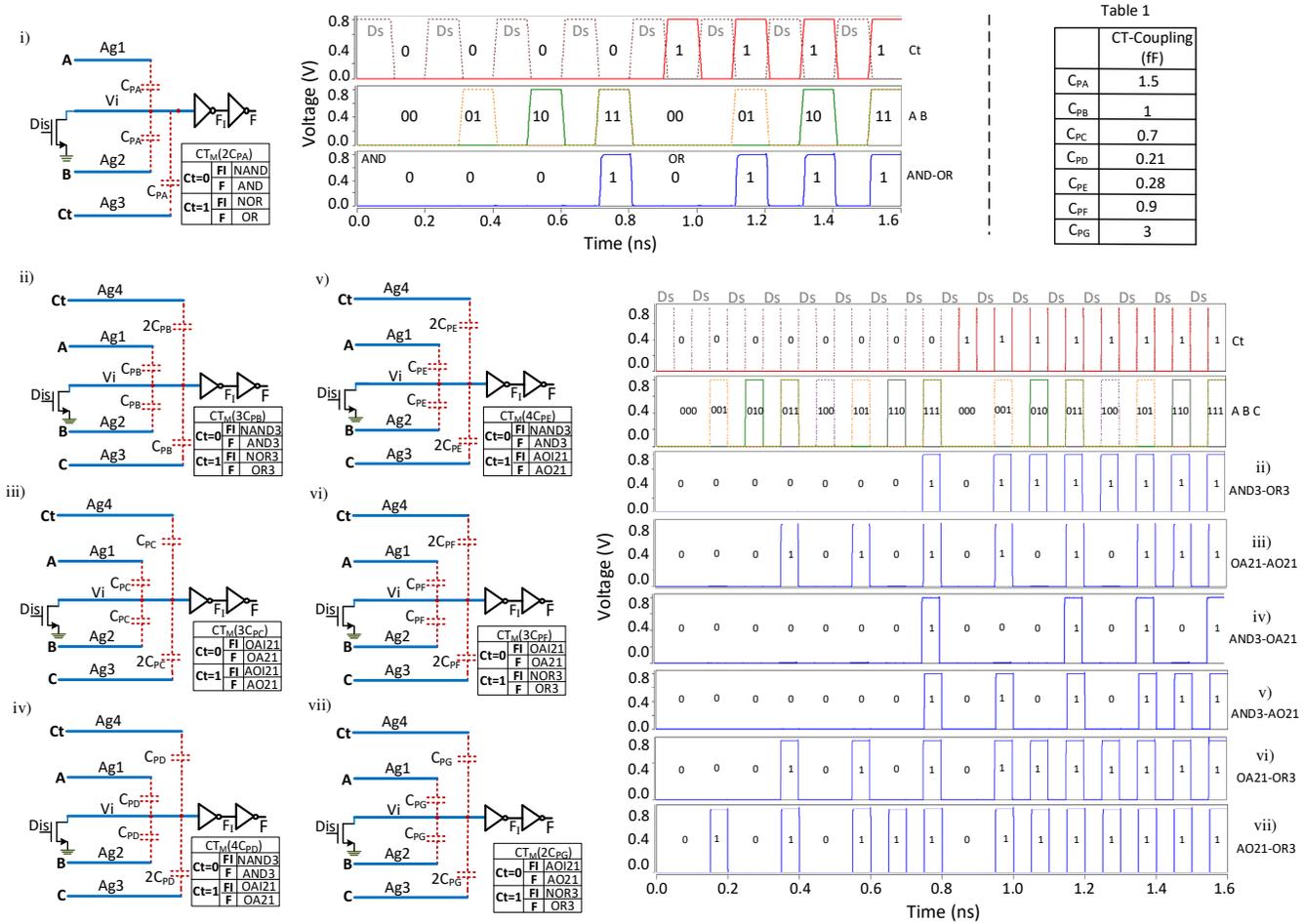

Fig.4. CT Polymorphic gates: i) AND2-OR2, ii) AND3-OR3, iii) AO21-OA21, iv) AND3-OA21, v) AND3 to AO21, vi) OR3-OA21, vii) OR3 to AO21.

The next six circuits depicted in the figures 4(ii) to 4(vii) implement 3 variable polymorphic functions. Therefore, in the response plots given adjacent, the input signals (*A*, *B*, *C*, *Dis* and *Ct*) are shown common, while panels below are the responses of the circuits from Fig.4(ii) to 4(vii). Fig.4(ii) depicts the 3 input AND-OR gate whose margin-function is $CT_M(3C_{PB})$, the three inputs (*A*, *B*, and *C*) are given $C_{PB}$ coupling, whereas, Ct aggressor is given twice the inputs, i.e., $2C_{PB}$. When control Ct=0 it operates as AND3, whereas, when Ct=1 the Ct aggressor (*Ag4*) augments charge through the coupling capacitance $2C_{PB}$, hence, following the function $CT_M(3C_{PB})$ the cell is now biased to operate as an OR3. The same response can be observed in the corresponding response graph (panel-3). The circuit responds as AND3 when *Ct=0* for first eight input combinations (000 to 111), whereas, it responds as OR3 when *Ct=1* during next eight combinations (000 to 111). Next, Fig.4(iii) shows the OA21-AO21 circuit which is a heterogenous-to-heterogeneous polymorphism. Here, aggressors *A*, *B*, and *Ct* are given $C_{PC}$ coupling, whereas input *C* is given $2C_{PC}$, the margin function is $CT_M(3C_{PC})$. When control *Ct=0* it operates as OA21, whereas, when *Ct=1* the Ct aggressor (*Ag4*) augments charge through the coupling capacitance $C_{PC}$, hence, following the function $CT_M(3C_{PC})$ the cell is now biased to operate as an AO21. The same response can be observed in the simulation graph (4th panel), the circuit responds as OA21 when *Ct=0* for first eight input combinations (000 to 111), whereas, it responds as AO21 when *Ct=1* for next eight combinations (000 to 111). Next, we show four different heterogeneous to homogeneous polymorphic circuits. Fig.4(iv) depicts the AND3-OA21 circuit, where, *A*, *B*, and *Ct* are given $C_{PD}$ coupling, while input *C* is given $2C_{PD}$, the margin function now is $CT_M(4C_{PD})$. When control *Ct=0* it operates as AND3, whereas, when *Ct=1* the Ct aggressor (*Ag4*) augments charge through the coupling capacitance $C_{PD}$, hence, following the function $CT_M(4C_{PD})$ the cell is now biased to operate as an OA21. The same response can be observed in the simulation graph (5th panel), the circuit responds as AND3 when *Ct=0* for first eight input combination (000 to 111), whereas, it responds as OA21 when *Ct=1* for next eight input combinations (000 to 111). Similarly, Fig.4(v) depicts AND3-AO21 circuit, where *A* and *B* are given $C_{PE}$ coupling, while *Ct* and *C* are now given

2$C_{PE}$ coupling, and the margin function here is $CT_M$ ($4C_{PE}$), therefore, the circuits respond (6th panel) as AND3 for all input combinations when *Ct=0*, whereas, it responds as AO21 when *Ct=1*. Similarly, Fig.4(vi) and Fig.4(vii) depict polymorphic OR3-OA21 and OR3-AO21 circuits respectively. The coupling choices for *A*, *B*, *C*, and *Ct* are as depicted in the circuit diagrams. The margin functions are $CT_M$ ($3C_{PF}$) and $CT_M$ ($2C_{PG}$) for OR3-OA21 and OR3-AO21 respectively. The simulation graphs in panel-7 and panel-8 show the response of corresponding circuits for all input combinations. When *Ct=0* for first 8 input combinations (000 to 111), the circuits in Fig.4(vi) and 4(vii) responds as OA21 and AO21 respectively, whereas, they both respond as OR3 when *Ct=0* for next 8 input combinations (000 to 111). It is worth noticing that, in all the cases, the control aggressor augments the charge (when it transitions from 0 to 1) required to bias the circuit to an alternate operation.

### A. CT-P Cascaded Circuit Example

To show the potential of CT polymorphic logic gates an example circuit of 2-bit multiplier-sorter (Fig.5) is implemented using the gates discussed above. The circuit uses 19 gates in total, 16 CT gates, and 3 inverters. 8 out of 16 CT gates are CT polymorphic gates. Polymorphic gates are efficiently employed to switch between the multiplier and sorter operations. A control signal (*Ct*) is used to switch between the operations, *Ct=0* is a multiplier and *Ct=1* is Sorter. Fig.6 shows the simulation response of the circuit, different operation modes of the circuit are annotated on top. They are Discharge State (DS), Multiplier (M) and Sorter (S). The first panel in the figure shows *Dis* and *Ct* signals, second and third panels show the two 2-bit inputs *A[1:0]* and *B[1:0]*, the following panels show the 4-bit response of the circuit *Y[3:0]*. To depict multiplier and sorter operations effectively, the *Ct* signal is given as 0 and 1 alternately which makes the circuit operate as multiplier and Sorter in successive logic states. Also, common inputs *A[1:0]* and *B[1:0]* are given for adjacent M and S modes. It can be observed from the response graphs (*Y[3:0]*) that, for same inputs, the circuit gives multiplier result when *Ct=0* and sorter result when *Ct=1*. For example, for the first input combinations, 10 and 11, the M operation gives 0110 as output and S operation gives 1110, similarly, for the second inputs, 01 and 01, M operation gives 0001 and S operation gives 1100. Similarly, M and S outputs are shown for few other combinations. The circuit consumes only 88 transistors. Thus CT-P circuits are compact, possess maximum reconfigurable features, and can efficiently implement larger polymorphic circuits in cascaded topology.

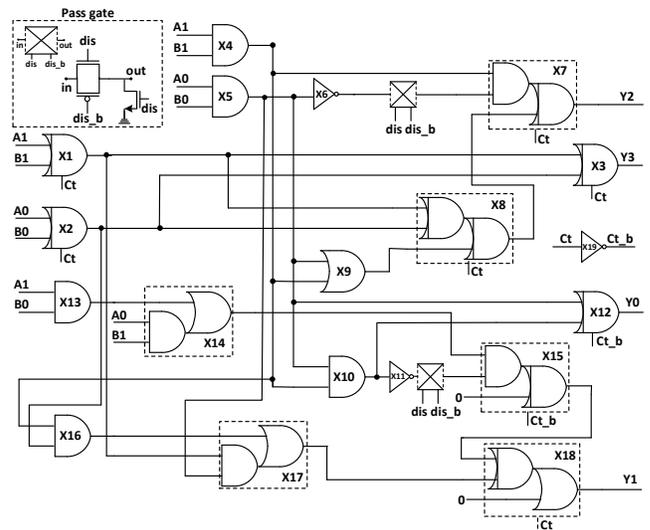

Fig.5. CT Polymorphic Multiplier-Sorter circuit

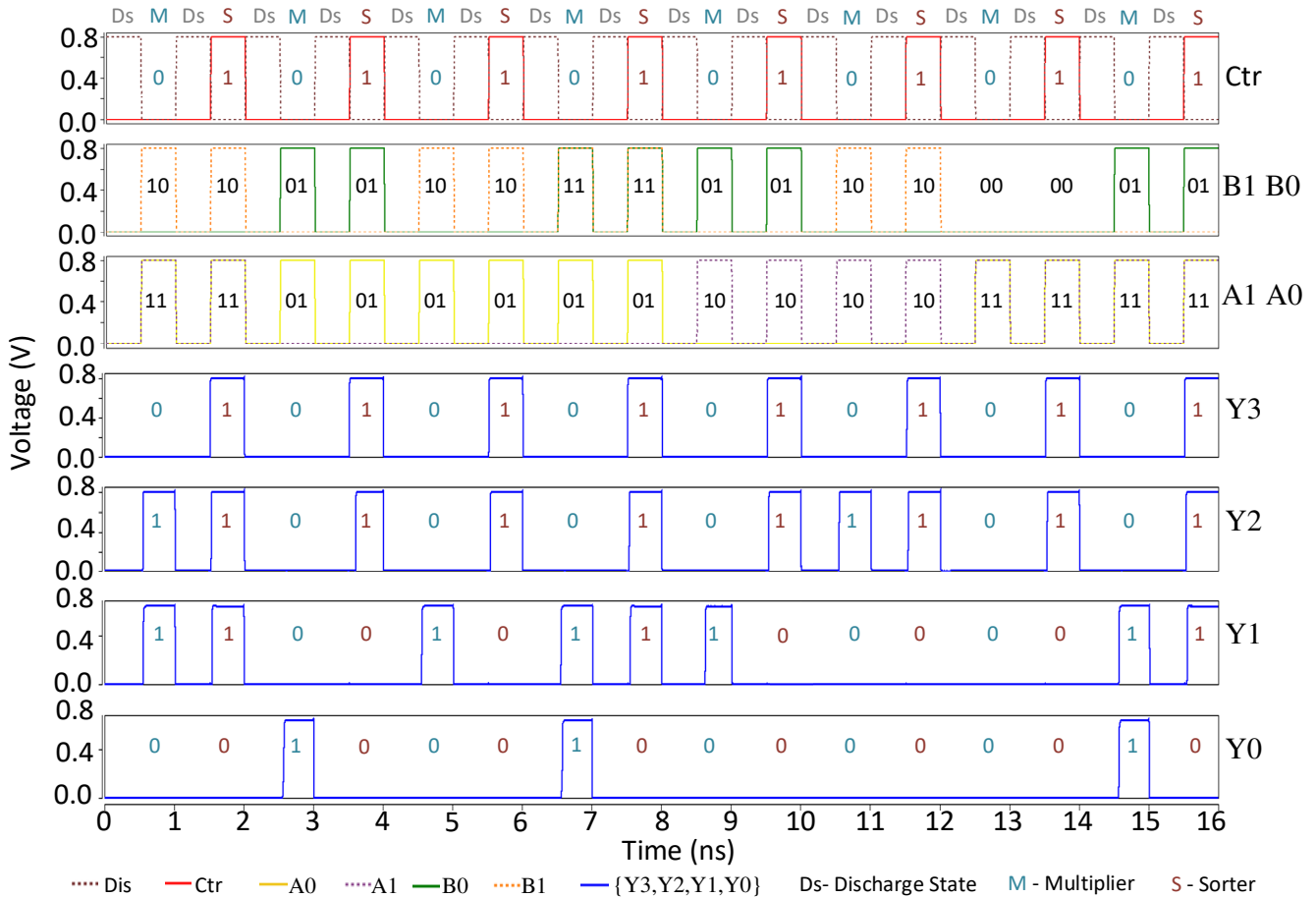

Fig.6. CT Polymorphic Multiplier-Sorter Simulation Results

### B. Discussion about Signal Integrity

As shown in [1], the actual computation in CT-logic happens in the nano-metal lines. However, to construct the larger circuits, the output voltage needs to be robust and possess enough drive-strength to drive the fan-out loads. These issues are addressed by connecting the victim node to an inverter. It acts as a regenerative Boolean threshold function, that is, it detects the logic levels computed on victim node and restore it to full swing (the victim voltages below the low logic threshold are restored to a logic high and voltages above the high logic threshold are restored to logic low). Nevertheless, this topology makes the CT-gates inverted logic functions (NAND, NOR etc.), for non-inverter logic functions (AND, OR etc.) an extra inverter is connected which also improves the signal further. This can be observed from the responses of inverting gates (Fig. 2 & 3) and non-inverting gates (Fig.4) wherein later case output signals are more robust. Also, it can be observed from Multiplier-Sorter results (Fig.6) that, the responses are robust in cascaded topology, and hence scalable to larger designs.

The other issue CT-cascaded topology faces is CT-logic specific monotonicity problem, which is, the CT-logic gates need the signal transition from 0 to 1 during each logic evaluation state for correct logic operation, thus if a logic high is retained on victim node from the previous operation it leads to logic failure. For example, when a CT-logic gate is driven by another inverting CT-logic gate (NAND, NOR etc..) it receives a logic high during DS from the prior gate, which would be carried to next evaluation stage leading to logic failure. This issue is resolved in this paper, by using a Pass-Gate (PG) solution (as depicted in inset figure in Fig.5), where, the inverting and non-inverting gate interfaces are connected through a transmission gate. The transmission gate passes the signal afresh during each evaluation state, and, similar to victim net, the corresponding aggressor net is discharged to ground in every DS (Fig.5). The other solution is by using a different set of CT-logic gates which operate on falling edge transition also, which are not presented here. Thus, a fully working large-scale compact polymorphic circuits, with reduced size, improved performance and power can be achieved using CT-logic style.

### IV. COMPARISON OF POLYMORPHIC TECHNOLOGIES

The crosstalk polymorphic (CT-P) logic technology is compared and benchmarked (Table.2) with respect to existing polymorphic approaches available in the literature. Different technology, device, and circuit metrics such as process node dependency, scalability, working mechanism, control parameter, performance trade-offs, and transistor count are compared and benchmarked. The CT-polymorphic approach compared to other approaches is a very compact implementation, friendly to advanced technology nodes and scalable to the larger polymorphic system. In addition, the working mechanism is simple and reliable. The benefits in performance metrics such as area, power, and performance are also best compared to any other approaches. Deliberate and very fast reconfigurability is achievable by using a control signal. The benchmarking of transistors count requirement for basic, complex and cascaded logic cases are given in the table. The complex gates listed for other approaches are constructed by cascading polymorphic NAND-NOR, AND-OR gates presented in [5-8]. The CT-P approach consumes fewer transistors than any other approach and moreover, a wide range of compact single-stage complex-function implementations like in CT-P was not reported in other approaches. The traditional approach ('CMOS' column in the table) is multiplexer based, where independent stand-alone circuits are designed and selected through a multiplexer. Though this approach is mainstream and can be implemented in any technology node (we have designed in 16nm), it consumes very large resources as listed in the table. Evolved circuits [3] are unconventional circuit structures evolved/synthesized using genetic algorithms. These circuits are strongly technology dependent (implemented .35um in [4]) and work only in special condition under which they are evolved, therefore, they are not adaptable to advanced technology nodes. Furthermore, they are inefficient in design; they suffer from unreliable responses (weak output logic level), lower input impedance, and high-power consumption etc. Hence, they are not scalable to larger designs and not usable as generic building blocks for digital polymorphic circuits. Next, to compare with emerging reconfigurable transistors we have considered ambipolar Si nanowire FET (SiNWFET) by De Marchi *et.al* [7]. In this approach, a nanowire transistor can be configured to either n-type or p-type with a control voltage. Limitations of this approach are, density benefit is limited, additional circuitry required to swap power rails for pull-up and pull-down networks, non-robust device response, and requirement of complex manufacturing steps. The other alternate approaches using emerging spintronic devices were also proposed[11]. However, they rely on complex information encoding scheme through spin-polarized currents and bipolar voltages. Consequently, they are a significant departure from existing computational device and circuit paradigms.

### V. CONCLUSION

We have discussed in this paper, a novel polymorphic logic fabric based on crosstalk-logic style. We have demonstrated polymorphic logic behavior between following functions, AND2-OR2, AND3-OR3, AO21-OA21, AND-AO21, AND3-

Table.2 Comparison of Polymorphic Technologies

| Technology | CMOS | Evolved Circuits[3] | | | Ambipolar NWFET[7] | Crosstalk-Polymorphic |
|---|---|---|---|---|---|---|
| Control parameter | Select Signal | Control Voltage | Temperature | Supply Voltage | Control voltage | Control Signal |
| Mechanism | Circuit duplication and use of multiplexers to select redundant blocks | A control voltage biases the circuits different operation | Temperature variation effects on devices bias the circuits to different modes | Power supply variation effects on devices biases the circuits to different mode | Band structure of the transistor is altered from p-type to n-type using a control gate | Signal Interference through interconnect crosstalk |
| Process-Technology Node | 16nm (independent) | 0.35um (strongly dependent) | | | 30nm (dependent) | 16nm (friendly to advanced technology nodes) |
| Scalability Dependence | Synthesis | Evolution limitation (Genetic Algorithms) | Evolution limitation (Genetic Algorithms) | Evolution limitation (Genetic Algorithms) | Large scale fabrication of nanowires and reliable ambipolar property | Crosstalk Couplings |
| Trade-off Vs. Custom ASIC | Density, power and performance penalties for redundant blocks | Power and performance penalties and limited density benefits | Power and performance penalties and limited density benefits | Power and performance penalties and limited density benefits | Limited density benefits | Density, Power and Performance benefits |
| **Transistor Count Comparison** | | | | | | |
| NAND-NOR | 14 | 11 (0/0.9) | 8 | 6(3.3/1.8) | 4 | 3 |
| AOI-OAI | 18 | 21 | 14 | 14 | 6 | 3 |
| AND2-OR2 | 18 | 10 (3.3/0) | 6(27/125$^0$C) | 8(1.2/3) | 6 | 5 |
| AND3-OR3 | 22 | 20 | 12 | 14 | 6 | 5 |
| AO21-OA21 | 22 | 21 | 12 | 16 | 8 | 5 |
| AND3-AO21 | 22 | 16 | 12 | 14 | 12 | 5 |
| AND3-OA21 | 22 | 16 | 12 | 14 | 12 | 5 |
| OR3-AO21 | 22 | 16 | 12 | 14 | 12 | 5 |
| OR3-OA21 | 22 | 16 | 12 | 14 | 12 | 5 |
| Multiplier-Sorter | 146 | 138 | | | 122 | 88 |

OA21, OR3-AO21, and OR3-OA21. A cascaded circuit example of multiplier-sorter is also presented. Our circuit evaluation and benchmark comparisons show that CT-P logic approach is very compact (i.e less device count) and efficient than any other polymorphic approach. The transistor count reduction with respect to different approaches ranges from 25% to 83%. For example, CT-P AOI21-OA21 cell shows 83%, 85% and 50% transistor count reduction, and multiplier-sorter circuit shows 40%, 36% and 28% transistor count reduction with respect to CMOS, genetically evolved and ambipolar transistor based polymorphic circuits respectively. Moreover, all CT-P logic gates are uniform, modular and generic in structure, and thus scalable to larger polymorphic digital systems.

REFERENCES

[1] Naveen Kumar Macha, *et al.,* "A New Concept for Computing Using Interconnect Crosstalks," Rebooting Computing (ICRC), 2017 IEEE International Conference, Washington, DC, USA, December 2017.
[2] McDermott, M.W., and Turner, J.E.: 'Configurable NAND/NOR element'. United States Patent 5,592,107, January 1997
[3] A. Stoica, R. Zebulum, and D. Keymeulen, "Polymorphic Electronics," Evolvable Syst. From Biol. to Hardw., vol. 2210, pp. 291–302, 2001.
[4] A. Stoica *et al.,* "Taking evolutionary circuit design from experimentation to implementation: some useful techniques and a silicon demonstration," in IEE Proceedings - Computers and Digital Techniques, vol. 151, no. 4, pp. 295-300, 18 July 2004.
[5] R. Ruzicka, "New Polymorphic NAND / XOR Gate 2 Known Polymorphic Gates," pp. 192–196, 2007.
[6] L. Sekanina, *et al.,*"Polymorphic gates in design and test of digital circuits," Int. J. Unconv. Comput., vol. 4, no. 2, pp. 125–142, 2008.
[7] M. De Marchi *et al.,* "Configurable logic gates using polarity controlled silicon nanowire gate-all-around FETs," IEEE Electron Device Lett., vol. 35, no. 8, pp. 880–882, 2014.
[8] J. Zhang, P. E. Gaillardon, and G. De Micheli, "Dual-threshold-voltage configurable circuits with three-independent-gate silicon nanowire FETs," Proc. - IEEE Int. Symp. Circuits Syst., pp. 2111–2114, 2013.
[9] Yu, W. J., Kim, U. J., Kang, B. R., Lee, I. H., Lee, E. H., Lee, Y. H.: Multifunctional logic circuit using ambipolar carbon nanotube transistor. Proc. SPIE 7399, 739906 (2009).
[10] Paasch, G., Lindner, Th., Rost-Bietsch, C.: Operation and Properties of Ambipolar Organic Field-effect Transistors, In: Journal of Applied Physics, Vol. 98, No. 8, 2005, US, pp. 084505-1 - 084505-13, ISSN 0021-8979, DOI 10.1063/1.2085314.
[11] S. Rakheja and N. Kani, "Polymorphic spintronic logic gates for hardware security primitives — Device design and performance benchmarking," 2017 IEEE/ACM International Symposium on Nanoscale Architectures (NANOARCH), Newport, RI, 2017, pp. 131-132.